\documentclass[12pt]{article}
\title{Symplectic approach to lightcone QCD}

\author{Alexey V. Popov\thanks{email: avp@novgorod.net} \\ \\ 
\small{\emph{Velikiy Novgorod, Russia}} 
}
\date{}

\begin{document}
\maketitle
\abstract{We develop a symplectic method of quantization of lightcone QCD.
We find that boundary gauge fields are crucial for a consistent and complete quantization. 
By applying the symplectic Faddeev-Jackiw method, we very carefully remove unphysical degrees of freedom and obtain the true phase space and the complete Hamiltonian.
The result is important for the high energy QCD evolution and for a further extension of the JIMWLK equation,
for which we find the second-order correction.
Finally, we make an important note about a peculiarity of four space-time dimensions. 
In additional, our method sheds new light on the lightcone quantization of a scalar field.
}
\newcommand{\ket}[1]{\ensuremath{|#1\rangle}}
\newcommand{\bra}[1]{\ensuremath{\langle#1|}}
\newcommand{\braket}[2]{\ensuremath{\langle#1|#2\rangle}}
\tableofcontents
\section{Introduction}
This paper is devoted to studying of a high energy scattering amplitude in QCD from \emph{ab initio}.
An important ingredient of high energy QCD is the Jalilian--Marian–-Iancu–-McLerran–-Weigert–-Leonidov–-Kovner (JIMWLK) equation \cite{CGC,JIMWLK}. 
Now, there exists the problem to make a step beyond the first-order calculations on which the JIMWLK equation is based.
The primary aim of the paper is to obtain the complete structure of quantum Hamiltonian at lightcone, 
which allows to study the all orders.
Recently, its importance was emphasized in Ref. \cite{Kovner07}
where it was used for the high-order studying of the high energy evolution in QCD.
Although in Ref. \cite{Kovner07} the right quantum Hamiltonian was used, the rigorous proof is not known. 
The long-standing problem of quantization is not solved yet.
In this paper we solve this problem and show explicitly how to obtain the complete Hamiltonian and the Hilbert space.

To quantize the theory we must identify the degrees of freedom. 
In lightcone theories, it is the well-known fact \cite{Steinhardt} that boundary conditions at $x^-=\infty$ are crucial for physical consequences. 
Namely, choosing a different boundary condition, we obtain a different physic.
Hence, in addition to the bulk fields, we must consider the boundary degrees of freedom.

There exist many fundamental works concerning the lightcone field theory \cite{Brodsky,Burkardt}. 
The problem of zero modes is also studied well \cite{Yamawaki}. 
However, we assert here that a complete solution is not given yet.
In a scalar theory, there is no natural boundary condition because such theories are too poor.
The main idea of Ref. \cite{Yamawaki} is to enumerate the whole set of boundary conditions and to take only conditions 
giving a consistent Poisson algebra with suitable quantization rules.
This way has two big problems.

The first big problem is that there is no natural choice of a boundary condition. 
Since a boundary condition is crucial for physical implications \cite{Steinhardt},
the theory becomes physically \emph{incomplete}.
It is known that, studying a lightcone theory, one must treat each type of boundary condition separately \cite{Devecchi}.
Thus, a boundary condition must be intrinsic property of a lightcone theory, not being chosen by hand.
In the QCD case, since the field $A_i$ is a gauge potential, we cannot simple set it to zero at the $x^-$-infinity. 
Hence, the important task is to find a physical assumption that allows us to impose a boundary condition on the field $A_i$ at the infinity.
The typical physical assumption is to select field configurations with finite energy \cite{Steinhardt,Zhang1}.
In the massive scalar case this easily gives the constraint $\varphi(\infty)=0$. So, boundary fields do no affect on the physical sector.
In the QCD, the situation is more complex.  
It was shown in Ref. \cite{Zhang1} that asymptotic of a gauge field contains a nontrivial and important information about the structure of QCD.
Unfortunately, in Refs. \cite{Zhang1,Zhang2} only the antisymmetric boundary condition is used. 
In this paper we prove that this is a wrong assumption.
Although this choice gives a consistent quantization, but it misses an important physical information about the theory.
In this paper we perform a more careful analysis and show how correctly to obtain boundary constraints and to quantize the theory. 

The problem of boundary conditions is closely related with the problem of zero modes
which causes the appreciable theoretical interest \cite{Brodsky,Yamawaki}. 
In is known that zero modes emerge only if the symmetric boundary condition is adopted \cite{Yamawaki}. 
In this paper we show that there are no zero modes in the lightcone QCD, 
since neither symmetric nor antisymmetric boundary conditions are applicable to the real QCD.
Fields at the infinity must be considered as independent variables.
This means that the method used in Ref. \cite{Brodsky} is inapplicable to the real world.
Instead, we show that nontrivial aspects of the theory are the complicated structure of the phase space
and the Hamiltonian containing the infinite number of terms.
So, our result disagrees with the Hamiltonian calculated in Ref. \cite{Brodsky} due to the different boundary condition.
Note that the results of Ref. \cite{Brodsky} is widely exploited in many works. 
As an example in the context of the high energy scattering, higher-order calculation was performed in Ref. \cite{Kovchegov} 
where running coupling corrections to the JIMWLK equation was calculated.

The second big problem is the inability of the conventional canonical method to treat fields at boundaries.
This problem is well known, the classical definition of the Poisson bracket fails, since the naive variational derivative does not exist. 
In Refs. \cite{Soloviev,Bering} it was proved that it is possible to generalize the Poisson bracket to a case with boundaries.
Although these generalizations seem challenging, currently there are no successful applications to physical problems. 
The situation becomes even more difficult if a theory has constraints.
Second class constraints are handled by a construction of the Dirac bracket which can be properly defined only in a finite-dimensional case.
In an infinite-dimensional case, due to the existence of surface integrals, there exists the problem to define the variational derivative.
The non-existence of the variational derivative was used in Ref. \cite{Steinhardt,Yamawaki} to conclude 
what a boundary condition gives an inconsistent theory.  
However, this is not the case.
Instead of the canonical method, in this paper we employ the symplectic Faddeev-Jackiw method \cite{Faddeev}
which allows to handle fields at the infinity. While naive variational derivative can be consistently defined only with 
zero boundary conditions, the symplectic method does not use it. 
A symplectic structure have no problems with the appearance of surface integrals.
Applying the symplectic method to QCD and not imposing any boundary condition, 
we remove unphysical degrees of freedom and obtain the phase space with the complete Hamiltonian.

The master logic of this paper is:
(1) To establish the physical importance of the gauge field at the $x^-$-infinities.
(2) To declare the necessity to quantize the theory without any boundary condition.
(3) To conclude that the traditional canonical method fails to do this. 
(4) To develop the symplectic method that can quantize the theory.
(5) Using the symplectic method, to quantize the theory.
(6) To check the consistence by the rederivation of the JIKWLK equation. 

The key result of the paper is the complete Hamiltonian of QCD.
The phase space of the theory is the space of fields $\tilde A_i(x^-,\vec x)$ 
that obey $\tilde A_i(+\infty,\vec x)=-\tilde A_i(-\infty,\vec x)$. The symplectic form is $\omega= \int \partial_- d\tilde A_i^a \wedge d\tilde A_i^a$.
The boundary field $\gamma_i^a(\vec x)$ is determined by the two following equations:
$$\partial_i  \gamma_j^a-\partial_j  \gamma_i^a+g f_{abc} \gamma^b_i \gamma^c_j=0$$
$$
\partial_i \gamma_i^a(\vec x)= \frac{g}{2}f_{abc}\gamma_i^b\gamma_i^c+ \int\limits_{-\infty}^{+\infty} \left(gf_{abc}\partial_-\tilde A_i^b \tilde A^c_i-gJ^+_a\right) dx^-
$$
where $J^+_a$ is an external current.
The original gauge fields $A_i$ is related to $\tilde A_i$ as 
$$
\tilde A_i(x^-,\vec x)= A_i(x^-,\vec x) - \frac{1}{2}\gamma_i(\vec x)
$$
The momentum $\pi^-_a(x^-,\vec x)$ is determined by condition $\pi^-_a(\pm\infty,\vec x)=0$ and the constraint 
$$
\partial_-\pi^-_a + \partial_- \partial_i A_i^a-gf_{abc}\partial_-A_i^bA^c_i+gJ^+_a=0
$$
The Hamiltonian is given by
$$
H[\tilde A_i]=\frac{1}{2} (\pi^-_a)^2+\frac{1}{4}F_{ij}^a[A_i] F^{ij}_a[A_i]  - g A_i^a J^i_a
$$

This paper is organized as follows.
In Sec. \ref{sect_3}, using the standard canonical method, we briefly review basic properties of the lightcone QCD.
In Sec. \ref{sect_1}, we formulate the symplectic Faddeev-Jackiw method. 
In Sec. \ref{sect_2}, we use this method to quantize a scalar field with the boundary condition which will be useful in the QCD case.
In Sec. \ref{sect_5}, we apply the symplectic method to the lightcone QCD. 
This section is divided on nine subsections:
(\ref{sub_1}) the primary analysis of the phase space, 
(\ref{sub_2}) the first step of the Faddeev-Jackiw algorithm and the complete set of Gauss' constraints,
(\ref{sub_3}) the analysis of the gauge invariance,
(\ref{sub_4}) the gauge fixing,
(\ref{sub_5}) the residual gauge transformations and the finite-energy condition,  
(\ref{sub_6}) the formulation of the complete Hamiltonian,
(\ref{sub_7}) the quantization of the theory and the construction of the Hilbert space,
(\ref{sub_8}) the perturbative expansion of the Hamiltonian operator,
(\ref{sub_9}) the third-order expansion which shows how the theory lives in space-times having more than four dimensions.
In Sec. \ref{sect_6} we use the first-order Hamiltonian to rederive the JIMWLK equation. Then we obtain the $O(\alpha_s^2)$-correction to this equation.
Section \ref{sect_4} contains our conclusions.

\section{QCD at lightcone.} \label{sect_3}
This section contains the standard canonical approach to QCD at lightcone. 
Also, we recall typical quantities and fix the notation.  

The covariant derivative is defined as 
\begin{equation}
D_\mu\psi=(\partial_\mu-igA_\mu) \psi
\end{equation}
where $\psi$ is a color multiplet.
The gauge curvature tensor is defined from the commutator of covariant derivatives
\begin{equation}
ig^{-1}[D_\mu,D_\nu] \psi= F_{\mu\nu} \psi
\end{equation}
\begin{equation} \label{eq_41}
F_{\mu\nu}=\partial_\mu  A_\nu-\partial_\nu  A_\mu-ig [A_\mu, A_\nu]
\end{equation}
In the subsequent calculations we shall sometimes use the the adjoint index notation 
\begin{equation}
A_\mu=A_\mu^a T_a
\end{equation}
\begin{equation}
[T_a,T_b]=if_{abc} T_c
\end{equation}
\begin{equation}
Sp(T_a T_b)=\frac{1}{2} \delta_{ab}
\end{equation}
\begin{equation}
F_{\mu\nu}^a=\partial_\mu  A_\nu^a-\partial_\nu  A_\mu^a+g f_{abc} A^b_\mu A^c_\nu
\end{equation}
The Lagrangian density of QCD is
\begin{equation} \label{eq_16}
L=-\frac{1}{4}F_{\mu\nu}^a F^{\mu\nu}_a + g A_\mu^a J^\mu_a
\end{equation}
where $J^\mu_a$ is a current of matter fields. For the fermion case it has a form
\begin{equation}
J^\mu_a=\bar \psi \gamma^\mu T_a \psi
\end{equation}
The equations of motion is
\begin{equation}
0=\frac{\delta L}{\delta A_\mu^a}=\frac{\partial L}{\partial A_\mu^a} - \partial_\nu \frac{\partial L}{\partial (\partial_\nu A_\mu^a) }
\end{equation}
Straightforward calculations give
\begin{equation}
\frac{\partial L}{\partial (\partial_\nu A_\mu^a) }=-F^{\nu\mu}_a
\end{equation}
\begin{equation}
\frac{\partial L}{\partial A_\mu^a}=gf_{abc}F^{\mu\nu}_bA^c_\nu + g J^\mu_a
\end{equation}
\begin{equation}
\frac{\delta L}{\delta A_\mu^a}=\partial_\nu F^{\nu\mu}_a + gf_{abc}F^{\mu\nu}_bA^c_\nu + g J^\mu_a
\end{equation}
A shorthand notation is
\begin{equation} \label{eq_2}
\frac{\delta L}{\delta A_\mu^a}=D_\nu F^{\nu\mu}_a+g J^\mu_a
\end{equation}
In the lightcone coordinates, the canonical momenta are
\begin{equation} \label{eq_5}
\pi^\mu_a= \frac{\partial L}{\partial (\partial_+ A_\mu^a) }=F^{\mu+}_a
\end{equation}
There is a primary first class constraint 
\begin{equation} \label{eq_4}
\pi^+_a=0
\end{equation}
Since $F^{i+}$ do not contain the velocities, the standard lightcone second class constraints are
\begin{equation} \label{eq_6}
\pi^i_a-F_a^{i+}=0
\end{equation}
where $i=1,2$. The Hamiltonian density is
\begin{equation} \label{eq_1}
H=\pi^\mu_a \partial_+ A_\mu^a-L
\end{equation}
The fundamental Poisson brackets are
\begin{equation}
\{A_\mu^a(x),\pi^\nu_b(y)\}=\delta_a^b \delta_\mu^\nu \delta(x-y)
\end{equation}
Since the Hamiltonian (\ref{eq_1}) is defined only on the constraint surface which is the image of the Legendre map,
an extension of the Hamiltonian on the whole phase space should be given. Usually, such extended Hamiltonian is called total Hamiltonian.
The consistency condition requires 
\begin{equation}
\{\pi^+_a, H\}=0
\end{equation} 
This leads to
\begin{equation} \label{eq_3}
0=\frac{\delta L}{\delta A_+^a}=D_\nu F^{\nu+}_a+g J^+_a=D_\nu\pi^\nu_a+gJ^+_a
\end{equation}
where we have used the definitions (\ref{eq_2}) and (\ref{eq_5}). Eq. (\ref{eq_3}) is the new first class constraint which is usually known as Gauss' law.

Now, we impose the first gauge fixing condition
\begin{equation}
A_+^a=0
\end{equation}
This converts the constraint (\ref{eq_4}) to a second class constraint. In this gauge we have
\begin{equation}
F_{+\mu}=\partial_+A_\mu
\end{equation}
\begin{equation}
\begin{array}{rl}H=&F^{\mu+} F_{+\mu}+\frac{1}{4}\left(2F_{+\mu} F^{+\mu}+2F_{-i} F^{-i}+F_{ij}F^{ij}\right)-gA_\mu J^\mu= \\
               & \frac{1}{2}F^{\mu+} F_{+\mu}+\frac{1}{2}F_{-i} F^{-i}+\frac{1}{4}F_{ij}F^{ij}-gA_\mu J^\mu = \\
               & \frac{1}{2}\left( F^{-+}F_{+-}+F^{i+}F_{+i} \right)+\frac{1}{2}F_{-i} F^{-i}+\frac{1}{4}F_{ij}F^{ij}-gA_\mu J^\mu = \\
  \end{array}
\end{equation}
Using the fact
\begin{equation}
F^{i+}=-F_{i-}
\end{equation}
we obtain the final result 
\begin{equation}\label{eq_9}
H=\frac{1}{2} \left( \pi^-\right)^2+\frac{1}{4}F_{ij}F^{ij}-gA_\mu J^\mu
\end{equation}
where we have dropped the obvious color summation index $a$. 

The second gauge fixing condition is
\begin{equation} \label{eq_7}
A_-^a=0
\end{equation}
This converts the constraint (\ref{eq_3}) to a second class constraint and the momenta $\pi^-$ becomes a depended variable.
The constraint (\ref{eq_6}) has been converted to
\begin{equation} \label{eq_10}
\pi^i-\partial_-A_i=0
\end{equation}
where we have used the identity $F^{i+}=F_{-i}$ and the gauge fixing (\ref{eq_7}). 
Similarly, in this gauge the constraint (\ref{eq_3}) can be expressed as
\begin{equation} \label{eq_8}
\partial_-\pi^-_a + \partial_- \partial_i A_i^a-gf_{abc}\partial_-A_i^bA^c_i+gJ^+_a=0
\end{equation}
The next task is to express $\pi^-$ from (\ref{eq_8}) and to substitute it into the Hamiltonian (\ref{eq_9}). 
Finally, one can obtain the Hamiltonian as a function of only the fields $A_i^a$ with the canonical structure 
derived from the constraints (\ref{eq_10}) via the standard Dirac brackets.

It is instructive to obtain the free version of the Hamiltonian by setting the coupling
constant to zero: $g=0$. Ignoring boundary terms, we have $(\pi^-)^2=(\partial_iA_i)^2$ and
this term is canceled by the equivalent term within $F_{ij}F^{ij}$. So, we have
\begin{equation} \label{eq_56}
H_0=\frac{1}{2}\partial_iA_j^a\partial_iA_j^a
\end{equation}
This Hamiltonian can be easily recognized as a sum of ordinary massless scalar field theories in the lightcone, which Hamiltonian
is $H_0=\frac{1}{2}(\partial_i \varphi)^2$. Hence, to quantize the free QCD, we can take 
the standard result of quantization of the free scalar theory. 
The index $j$ in (\ref{eq_56}) directly corresponds to transverse polarization of a physical gluon.

\newcommand{\Ker}{\mathop{\rm Ker}\nolimits}
\newcommand{\im}{\mathop{\rm Im}\nolimits}
\section{Faddeev-Jackiw method} \label{sect_1}
In this section we briefly review a simplified and geometrized version of the symplectic Faddeev-Jackiw method \cite{Faddeev}. 
The method is a useful tool to study constrained dynamical systems.  

Consider a mechanical system whose Lagrangian is 
\begin{equation} \label{eq_21}
L(q_i,\dot q_i)=a_i(q)\dot q^i - V(q)
\end{equation} 
Since it is linear over the velocities $\dot q^i$, we have a constrained system.
This system can be studied by both the standard canonical method and the original Faddeev-Jackiw method \cite{Faddeev}.
However, if one needs observables that depend only on the coordinates $q^i$,
we can directly endow the coordinate manifold $Q$ with a symplectic form.
A Lagrangian $L(q_i,\dot q_i)$ can be viewed as a function on a tangent bundle.
In each fiber, the expression $a_i(q)\dot q^i$ of the Lagrangian (\ref{eq_21}) defines a linear form. 
Hence, on the whole manifold we have the differential 1-form  $\xi=a_i(q) dq^i$.
The exterior differential $d$ produces 2-form
\begin{equation}
\omega=d \xi=\frac{\partial a_i}{\partial q^j} dq^j \wedge dq^i = 
\frac{1}{2}\left(\frac{\partial a_i}{\partial q^j}-\frac{\partial a_j}{\partial q^i}\right)  dq^j \wedge dq^i
\end{equation}
where $\wedge$ is the wedge product. The form $\omega$ is closed: $d\omega=0$ due to nilpotency $d^2=0$.
To be a symplectic from, a closed 2-from must be nondegenerate. 
If $\omega$ is nondegenerate, the phase space is completely specified
and a Poisson bracket is generated by the inverse of the skew-symmetric matrix $\omega_{ij}$
\begin{equation}
\{f_1,f_2\}=(\omega^{-1})^{ij}\frac{\partial f_1}{\partial q^i}\frac{\partial f_2}{\partial q^j}
\end{equation}
The Hamiltonian is simple 
\begin{equation}
H=V(q)
\end{equation}
If $\omega$ is degenerate we must make the additional steps. 
A degenerate 2-form $\omega$ at each point of the manifold defines its own kernel in the linear tangent space. 
For any vector $a \in \Ker \omega$ and for any tangent vector $b$, we have $\omega(a,b)=0$.
It is easy to check that the Lie bracket of two vector fields $a,b\in\Ker \omega$ is also an element of $\Ker \omega$. 
This fact is a consequence of $d\omega=0$ and the Cartan formula for $d\omega(a,b,c)$ where $a,b\in \Ker \omega$.
Hence, due to Frobenius theorem, $\Ker \omega$ is an integrable set of vectors and it at each point defines a submanifold. 
So, if there are no topological obstructions, we have a \emph{foliation} of the original coordinate manifold
which becomes a union of parallel submanifolds of smaller dimension.
A fiber of the foliation will be also called $\Ker \omega$.
The foliation gives an equivalence relation: 
two points are equivalent to each other if they belong to a same submanifold.
The main statement is: there exists a manifold $Q_1$ and a natural isomorphism $i$ such that
\begin{equation} \label{eq_22}
Q_1 \times \Ker \omega \stackrel{i}{=} Q
\end{equation}
Also, the isomorphism $i$ induces a closed and nondegenerate 2-form $\omega_1$ on the manifold $Q_1$, since $\Ker \omega$ is factored out.
Hence, the form $\omega_1$ is a symplectic form on $Q_1$.
For practical purposes, instead of a usage of the formal statement (\ref{eq_22}) it is sufficient to find coordinates that 
explicitly separate $\Ker \omega$ from $Q$. 
Let $\{z_i\}$ be coordinates in $\Ker \omega$ and $\{\tilde q_k\}$ be the remaining coordinates. 
In this coordinates the form $\omega$ does not depend on the forms $dz_i$
\begin{equation} \label{eq_29}
\omega=\tilde \omega_{kl}(\tilde q) d\tilde q^k \wedge d\tilde q^l
\end{equation}
Also, it should be stressed that $\tilde \omega_{kl}(\tilde q)$ cannot depend on the variables $z^i$ due to the property $d\omega=0$.
So, the variables $\tilde q^k$ form the manifold $Q_1$ endowed with the symplectic form $\omega_1=\tilde \omega(\tilde q)$.
Note that in the symplectic analysis the decomposition (\ref{eq_22}) and the well-known Darboux's theorem have the same origin.

Since we have $Q_1=Q/\Ker \omega$, the manifold $Q_1$ is a quotient manifold, while the Hamiltonian $H$ is a function on $Q$.
Hence, in general cases $H$ cannot be reduced on $Q_1$.
This is the most intimate point of the story.
To obtain a reduced Hamiltonian, we can try to fix a natural left inverse of the canonical epimorphism $Q \to Q_1$.
The fixation is based on the fiber diffeomorphisms whose vector fields lie in $\Ker \omega$.
Namely, we construct a condition
\begin{equation} \label{eq_30}
dH(\Ker \omega)=H(\tilde q,z+\delta z)-H(\tilde q,z)=0
\end{equation}
In the coordinate notation, we are looking solutions of the equation 
\begin{equation} \label{eq_23}
\frac{\partial H(\tilde q,z)}{\partial z^i}=0
\end{equation}
Taking coordinates $\tilde q^k$ of a point in $Q_1$, we solve Eq. (\ref{eq_23}) and find the remaining one $z^i$.
The reduced Hamiltonian will be $H_1(\tilde q)=H(\tilde q, z(\tilde q))$.
Nevertheless, we have problems again.
Since, in general, Eq. (\ref{eq_23}) can have: no solutions, precisely one solution, more than one solution, and any $z^i$ as a solution.
Since the last alternative arises if $H$ does not depend on $z^i$, we can restrict $H$ on $Q_1$ trivially.
If there are no solutions, we must remove such points $\tilde q^k$ from the theory. Hence, we have new constraints.
More precisely, if Eq.~(\ref{eq_23}) has a solution only in a submanifold $Q_2 \subset Q_1$, then $Q_2$ contains the true phase space of the theory.
Currently, we are not concerning the case of more than one solution, 
 since we shall see that in the application to QCD only the first and last alternatives are required.
 
Finally, after the solution of Eq.~(\ref{eq_23}) we have a submanifold $Q_2$ with Hamiltonian $H_2\equiv H_1$.
Again, if $Q_2\neq Q_1$ we have a problem. The reduced on $Q_2$ symplectic form $w_1$ may be degenerate. We denote it $\omega_2$.
If $\omega_2$ is a degenerate 2-form in $Q_2$, then, to remove $\Ker \omega_2$, we just repeat the procedure described above.
This gives new pairs $(Q_3,\omega_3)$, $(Q_4,\omega_4)$ and so on, until at some step $n$ we shall find the true phase space
that is a triple: a coordinate manifold $Q_n$, a nondegenerate closed 2-form $\omega_n$, and a Hamiltonian $H_n$ that is a function on $Q_n$.
Now, the method is completely specified.

It is quite remarkable that the presented method can be safely applied to systems having infinite and continuous number of degrees of freedom,
because functional derivatives are not required.
There is only one potentially problem to invert a symplectic form, but it can arise only at the last step of the construction.

The method is also useful to impose an external constraint.
Since a differential form can be naturally restricted on a submanifold
and the Hamiltonian is 0-form,  
we are able to impose by hand any constraint which defines a submanifold in the original manifold.
A dynamical system will be completely specified if a reduced symplectic form is nondegenerate.
How to consider  the degenerate case it has been described above in this section.
External constraints are required to impose a gauge fixing condition.
Since there is freedom to choose a gauge, in Sect. \ref{sect_5} we shall use a manually constructed external constraint.

\section{New lightcone quantization of a scalar field} \label{sect_2}
In order to understand the role of fields at the boundary we consider a scalar field theory in $1+1$ dimensions. 
The free Lagrangian is
\begin{equation}
L=\int \partial_- \varphi \partial_+\varphi dx^-
\end{equation}
Since $L$ is linear over the velocities $\partial_+\varphi$,  
we are able to apply simplified Faddeev-Jackiw method which was constructed in Sec. \ref{sect_1}.
Since the coordinate space is a linear space, the tangent space is naturally isomorphic to it. 
The Lagrangian $L$ induces 1-form 
\begin{equation}
\xi=\int \partial_- \varphi(x) d\varphi(x) dx
\end{equation}
where $d\varphi(x)$ is a functional analog of 1-forms.
It acts on a vector $u(x)$ as $d\varphi(x) \left(u\right)=u(x)$.
Also, $L$ induces a symplectic 2-form $\omega$ in the linear space of all fields $\varphi$ 
\begin{equation} \label{eq_28}
\omega=d\xi=\int \partial_-d\varphi(x) \wedge d\varphi(x) dx
\end{equation}
where $\wedge$ is the continuous wedge product.
Taking any two fields $\varphi_1(x^-)$ and $\varphi_2(x^-)$, the form $\omega$ can be expressed as
\begin{equation} \label{eq_13}
\omega(\varphi_1,\varphi_2)=\int \left( \partial \varphi_1 \varphi_2 -  \varphi_1 \partial\varphi_2\right) dx^-
\end{equation}
where we have used a shorthand notation $\partial\equiv\partial_-$. 
The constant symplectic form $\omega$ promotes the space of field to a phase space. 
Note that if there are no restrictions on the field values at the infinity, the form $\omega$ is not degenerated
because one cannot find a field $\varphi_1$ such that the action of the form (\ref{eq_13}) is zero at any $\varphi_2$.
Only if we impose the symmetric boundary condition $\varphi(-\infty)=\varphi(+\infty)$, then $\omega$ degenerates
in the space of field configuration that is symmetric at infinity because $\omega(const,\varphi)=0$ for any $\varphi$.
In this case there exists a so-called zero mode -- the constant field which commute with each other field.  
This is the typical result of Discretized LightCone Quantization (DLCQ) with the symmetric boundary condition. 
Nevertheless, in general, the symmetric boundary condition can have physical meaning only in a nontrivial topology case. 

With aim to apply the result to QCD we impose the following boundary condition
\begin{equation} \label{eq_11}
\varphi(-\infty)=0
\end{equation}
and there are no restrictions on $\varphi(+\infty)$.
The condition (\ref{eq_11}) will arise in QCD due to the opportunity to impose a gauge fixing condition.

To obtain a Poisson bracket we have to find an inverse of the matrix $\omega$.
However, a direct inversion meets with math difficulties. 
A careful inversion requires some math stuff which will be developed now.

Let us define three linear spaces. 
Let $Q$ be the space of all fields $\varphi(x^-)$. 
Let $Q_0$ be the space of fields obeying the constraint (\ref{eq_11}). 
Let $\tilde Q$ be the space of fields obeying the antisymmetric boundary condition $\varphi(-\infty)+\varphi(+\infty)=0$.
There is an important fact that $Q_0$ and $\tilde Q$ are linear subspaces of $Q$. 
This means that these subspaces are invariant under any linear combination $a\varphi_1+b\varphi_2$.
As an example, constraints like $\varphi(-\infty)=1$ do not generate linear subspaces.
The symplectic form $\omega$ can be naturally restricted to a linear subspace.
In general, a restricted form can be degenerated, 
 but in our case it is easy to check that the corresponding forms in $Q_0$ and $\tilde Q$ are not degenerated. 
Hence, the spaces $Q_0$ and $\tilde Q$ have a natural phase space structure.
The key step of our setup is a construction of a linear map $\chi: Q_0 \to \tilde Q$ 
\begin{equation} \label{eq_12}
\chi(\varphi_0)=\varphi_0-\frac{1}{2}\varphi_0(+\infty)
\end{equation}
where $\varphi_0\in Q_0$.
It is clear that $\Ker\chi=0$ and $\im \chi=\tilde Q$. 
Moreover, there exists the unique inverse map $\chi^{-1}: \tilde Q \to Q_0$
\begin{equation} \label{eq_14}
\chi^{-1}(\tilde \varphi)=\tilde \varphi+\tilde \varphi(+\infty)
\end{equation}
where $\tilde \varphi \in \tilde Q$. Hence, we have proved that the map $\chi$ is an \emph{isomorphism} of vector spaces. 
This fact allows us to use $\chi$ as a \emph{change of variables} in the dynamic system.
Hence, we shall use the space $\tilde Q$ as new coordinates of the system.
Also, the map $\chi$ is a \emph{symplectic map} (symplectomorphysm). This means that for any $\varphi,\psi \in Q_0$ we have
\begin{equation}
\omega(\chi(\varphi),\chi(\psi))-\omega(\varphi,\psi)=
\frac{1}{2}\varphi(\infty)\int\ \partial \psi - \frac{1}{2}\psi(\infty)\int \partial \varphi=0
\end{equation}
So, in the sense of Hamiltonian mechanics, $\chi$ is a \emph{canonical transformation}.

At first sight, the map (\ref{eq_12}) seems strange. 
To see it as a linear map we replace the whole $x^-$-line $[-\infty,+\infty]$ by a finite interval $[-L,+L]$ 
and approximate by a finite lattice $\varphi_i=\varphi(x_i)$, $i=0\ldots n$, $x_0=-L$, $x_n=+L$. 
The finite-dimensional linear space formed by the set $\{\varphi_i\}$ is an approximation to the space $Q$.
The map (\ref{eq_12}) becomes $\tilde \varphi_i=\varphi_i-\varphi_0/2$. 
In the matrix notation we have $\tilde \varphi_j=(\delta_{ji}-\delta_{0i}/2) \varphi_i$.
Thus, we see that the map (\ref{eq_12}) can be approximated by a linear matrix $C_{ji}=\delta_{ji}-\delta_{0i}/2$.

Having the space $\tilde Q$ and the symplectic form $\omega$ (\ref{eq_13}), we are ready to find a Poisson structure. 
This can be achieved by the search of a linear operator $P(x-y)$ such that 
\begin{equation} \label{eq_15}
\omega\left(\varphi_1,\int P(x-y) \varphi_2(y) dy\right)=\int \varphi_1 \varphi_2 dx
\end{equation} 
for $\varphi_1,\varphi_2 \in \tilde Q$.
For brevity, we just show here that a standard ansatz $P(x)=a\varepsilon(x)$ 
 motivated by the well-known results of lightcone field theories solves the task.
Here we have used a sign function
\begin{equation}
\varepsilon(x)=\left\{ \begin{array}{l} \phantom{-}1 \quad \mbox{if }x>0\\\phantom{-}0 \quad \mbox{if }x=0\\-1 \quad \mbox{if }x<0 \end{array} \right.
\end{equation} 
Its derivative is $\varepsilon'(x)=2\delta(x)$.
Using the rule 
\begin{equation}
\int\limits_{-\infty}^{+\infty} \partial \tilde \varphi(x) \varepsilon(x-y)dx=
\int\limits_y^{+\infty} \partial \tilde\varphi(x) dx-\int\limits_{-\infty}^y \partial \tilde\varphi(x) dx=-2\tilde\varphi(y)
\end{equation}
we find the solution of Eq. (\ref{eq_15})
\begin{equation}
P(x-y)=-\frac{1}{4}\varepsilon(x-y)
\end{equation}
The final step is a quantization of the theory. This is the standard textbook result so we write only the final result for completeness.
Restoring the transverse coordinates, we declare the quantization rules  
\begin{equation} \label{eq_54}
\left[\hat \varphi(x^-,\vec x),\hat \varphi(y^-,\vec y)\right]=-\frac{i}{4}\varepsilon(x^--y^-) \delta(\vec x-\vec y)
\end{equation}
\begin{equation} \label{eq_68}
\hat \varphi(x^-,\vec x)=
 \int\limits_{k^+>0}\frac{1}{\sqrt{2k^+}}\left(
         \hat{a}^\dag_k e^{ik^+x^--i\vec{k}\vec{x}}+ 
         \hat{a}_k e^{-ik^+x^-+i\vec{k}\vec{x}}
         \right) \frac{d^2k dk^+}{\sqrt{(2\pi)^3}}
\end{equation}
\begin{equation} \label{eq_55}
\left[ \hat a_k^{\phantom{\dag}}, \hat a^\dag_p\right]=\delta(k^--p^-) \delta(\vec k-\vec p)
\end{equation}
where the restriction $k^+>0$ is imposed since the free Hamiltonian must be bounded from below. 

A price for using the map (\ref{eq_12}) is that the Hamiltonian should be recalculated in terms of fields $\tilde \varphi$.
Using the inverse map (\ref{eq_14}), we have $H=H(\tilde \varphi+\tilde \varphi(+\infty))$.
This is a potentially dangerous expression, since we must define a meaning of the operator $\hat\varphi(+\infty)$ and 
to check commutation relations between observables and the Hamiltonian.
In Ref. \cite{Kovner07} a similar operator is called zero mode.
However, $\varphi(+\infty)$ is not a zero mode. 
It is just a tool to calculate right expressions for observables such as a Hamiltonian, preserving friendly commutation relations.

\section{Complete Hamiltonian of QCD at lightcone} \label{sect_5}
In this section we apply the Faddeev-Jackiw method \cite{Faddeev} to QCD.
\subsection{Bare phase space} \label{sub_1}
Bare phase space is the space that one obtains after the first step of the Faddeev-Jackiw algorithm.
Separating the velocities from the QCD Lagrangian density (\ref{eq_16}) which at lightcone has the form
\begin{equation}
L=\frac{1}{2}{F_{+-}}^2+F_{+i}F_{-i}-\frac{1}{4}F_{ij}F^{ij}+gA_\mu J^\mu
\end{equation}
we identify the following schematic structure 
\begin{equation}
L=L_1+ L_2 \partial_+ A_i + L_3 \partial_+A_- +  \frac{1}{2} (\partial_+A_-)^2  
\end{equation}
where $L_i$ do not depend on velocities. 
There is no dependence on $\partial_+ A_+$, there is a linear dependence on $\partial_+ A_i$, and
there is a typical quadratic dependence on $\partial_+A_-$.
So, the simplified Faddeev-Jackiw method cannot be directly applied.
To obtain a phase space and a symplectic structure we have to perform the Legendre map only over the variable $\partial_+A_-$
\begin{equation}
H=\pi^- \partial_+ A_--L
\end{equation}
where the terms, having the velocities $\partial_+ A_i$, have been removed into the canonical 1-form by the way which was shown in Sec. \ref{sect_1},
and the velocities $\partial_+ A_-$ are extracted from the identity 
\begin{equation}
\pi^-=F_{+-}
\end{equation}
So, the Hamiltonian density is
\begin{equation} \label{eq_17}
H=\frac{1}{2} (\pi^-_a)^2+\pi^-_a\left( \partial_- A_+^a - g f_{abc} A^b_+ A^c_- \right)
  +\left(\partial_iA_+^a - gf_{abc}A_+^bA_i^c \right) F_{-i}^a+\frac{1}{4}F_{ij}^a F^{ij}_a - g A_\mu^a J^\mu_a
\end{equation}
The linear phase space is the set of fields 
\begin{equation} \label{eq_18}
\{A_+,A_-,\pi^-,A_i\}
\end{equation}
The canonical 1-form is
\begin{equation}
\xi =\int \pi^-_a dA_-^a + \int F_{-i}^a dA_i^a
\end{equation}
where the integration over the space coordinates $x^-,x^i$ is assumed. The canonical 2-form is
\begin{equation} \label{eq_25}
\omega= d \xi = \int d\pi^-_a \wedge dA_-^a + \int dF_{-i}^a \wedge dA_i^a
\end{equation}
Note that $\omega$ is a degenerate form since, at least, it does not contain $dA_+$. 

\subsection{The complete set of Gauss' constraints} \label{sub_2}
We have seen that the 2-form (\ref{eq_25}) does not depend on $dA_+$ and $A_+$. 
Hence, the choice of coordinates (\ref{eq_18}) explicitly separates $\Ker \omega$ and 
Eq. (\ref{eq_25}) already has the form of Eq. (\ref{eq_29}). 
So, the coordinate $A_+$ plays the role of variables $z^i$ of Sec. \ref{sect_1}.
To obtain the reduced Hamiltonian, we have to impose a functional extension of the condition (\ref{eq_30}).
Making a variation $\delta A_+$, we have 
\begin{equation} \label{eq_31}
0=\int(H[A_++\delta A_+]-H[A_+])d^3 x
\end{equation}
For fixed $A_-,\pi^-,A_i$, Eq. (\ref{eq_31}) has no solutions or any $A_+$ as a solution. 
Hence, in accordance with the discussion in Sec. \ref{sect_1}, we have a constraint within the bare phase space.
The bulk part of the variation (\ref{eq_31}) gives the ordinary Gauss' law (\ref{eq_3})
\begin{equation} \label{eq_35}
D_-\pi^-_a+D_iF_{-i}^a+gJ^+_a=0
\end{equation}
while the boundary one gives two new constraints
\begin{equation} \label{eq_32}
\pi^-\left({x^-=\pm\infty},\vec x\right)=0
\end{equation}
\begin{equation} \label{eq_33}
F_{-i}\left(x^-,{|\vec x|=\infty}\right)=0
\end{equation}
and the reduced Hamiltonian density is
\begin{equation} \label{eq_34}
H=\frac{1}{2} (\pi^-_a)^2+\frac{1}{4}F_{ij}^a F^{ij}_a - g A_-^a J^-_a - g A_i^a J^i_a
\end{equation}

The constraints (\ref{eq_32}) and (\ref{eq_33}) are the necessary and important part of the theory.
In particular, we shall see below that the constraint (\ref{eq_32}) is very helpful in the construction of the true phase space.

The constraints (\ref{eq_35}) and (\ref{eq_32}) completely determine the coordinate $\pi^-$. 
It will be useful to find $\pi^-$ and to substitute it into the Hamiltonian (\ref{eq_34}).
However, one should be careful, since we shall see that the constraints (\ref{eq_35}),(\ref{eq_32}) also restrict 
the coordinates $A_i$.

\subsection{Gauge invariance} \label{sub_3}
Until now, we did not use any gauge fixing procedure. 
In principle, we can avoid this step by the introduction of unphysical degrees of freedom into the scene. 
Nevertheless, when using noncovariant methods, a gauge fixing allows to simplify a calculation.
How to motivate a gauge fixing within the context of the symplectic method?
In the canonical approach, the Dirac procedure says that in a presence of first class constraints 
we can impose new gauge fixing constraints that convert all constraint to the second class. 
In the symplectic method a gauge invariance can be handled in a more straightforward way.
Let us recall the definition of a gauge transformation. It is a matrix-valued function $U(x^-,x^i)\in SU(N)$.
In the matrix notation the gauge field $A_\mu$ transforms as
\begin{equation} \label{eq_19}
\tilde A_\mu=U A_\mu U^\dag - \frac{i}{g} (\partial_\mu U) U^\dag
\end{equation}
The Hamiltonian (\ref{eq_17}) is invariant under $U$ by the intrinsic gauge invariance of the theory. 
The Gauss' constraints (\ref{eq_35}), (\ref{eq_32}), and (\ref{eq_33}) is explicitly invariant too, 
since they transform covariantly. 

Since we have used the differential calculus, it is instructive to review a some geometric stuff.
The primary fields $\{A_+,A_-,A_i\}$ at fixed $x^+$ forms an infinite-dimensional manifold. 
The gauge transformation (\ref{eq_19}) acts like a diffeomorphism. 
Since a diffeomorphism can be naturally extended on the tangent bundle, 
any gauge transformation has the following action on a tangent vector:
\begin{equation} \label{eq_20}
\tilde {\dot A_\mu}=U \dot A_\mu U^\dag
\end{equation}
where the function $\dot A_\mu$ implements a tangent vector.
Eq. (\ref{eq_20}) is a consequence of the well-known fact that a difference between two connections is an adjoint vector under a gauge transformation.
Also, the transformation (\ref{eq_20}) naturally induces the action of the gauge group on a cotangent vector. 
Namely, the differentials $dA$ transforms as 
\begin{equation} \label{eq_24}
\tilde {dA_\mu} = U^\dag dA_\mu U
\end{equation}
Eq. (\ref{eq_24}) is a consequence of the requirement that the convolution $dA_\mu(A_\nu)$ must be a gauge invariant.
Similarly, since $\pi^-$ is an element of the cotangent bundle, it transforms in according to (\ref{eq_24}).
So, we can conclude that the 2-form (\ref{eq_25}) is a gauge invariant.

\subsection{Main gauge fixing} \label{sub_4}
Since the Hamiltonian, the Gauss' constraints, all physical observables, and the canonical form $\omega$ are gauge invariants, 
the gauge invariance (\ref{eq_19}) is a genuine invariance of the dynamical system.
So, we have the opportunity to impose a constraint on the set of fields (\ref{eq_18}) within the bare space $Q$.
Mathematically, this is just a choice of a submanifold $Q_1 \subset Q$. 
Certainly, we must prove that such a choice is physically admissible. 

We use the standard lightcone gauge
\begin{equation} \label{eq_26}
A_-^a=0
\end{equation}
It is easy to show that each field configuration has a gauge equivalent obeying (\ref{eq_26}) and there are no Gribov ambiguities. 
Hence, the choice (\ref{eq_26}) does not restrict the physical sector of the theory. 
A gauge transformation that leads to the condition (\ref{eq_26}) is the usual parallel transporter  
\begin{equation}
U(x^-,\vec x)=P \exp \left(-ig \int_{x^-_0}^{x^-} A_-(x^-,\vec x) dx^-\right)
\end{equation}
where the coordinate $x^-_0$ of the initial plane is arbitrary.

After the gauge fixing, the canonical 2-form (\ref{eq_25}) becomes
\begin{equation} \label{eq_38}
\omega= \int dF_{-i}^a \wedge dA_i^a = \int \partial_- dA_i^a \wedge dA_i^a
\end{equation}
which is just a vector version of the form (\ref{eq_28}) in a scalar theory.
The Gauss' constraint (\ref{eq_35}) becomes 
\begin{equation} \label{eq_36}
\partial_-\pi^-_a + \partial_- \partial_i A_i^a-gf_{abc}\partial_-A_i^bA^c_i+gJ^+_a=0
\end{equation}

Now, after the gauge fixing and the elimination of $\pi^-$, our phase space 
is entirely specified by the set of the transverse gauge potentials $A_i$ obeying the constraint (\ref{eq_33}).
This space is endowed with the closed 2-form (\ref{eq_38}) and the Hamiltonian
\begin{equation} \label{eq_40}
H=\frac{1}{2} (\pi^-_a)^2+\frac{1}{4}F_{ij}^a F^{ij}_a  - g A_i^a J^i_a
\end{equation}

We mentioned above that the Gauss' constraints also impose additional constraints on the variables $A_i$.
To see the hidden set of constraints, we integrate the Gauss' constraint (\ref{eq_36}) over the space variable $x^-$.
Using Eq. (\ref{eq_32}), we have
\begin{equation} \label{eq_37}
\partial_i A_i(+\infty,\vec x) =  \partial_i A_i(-\infty,\vec x) + G[A_i,J]
\end{equation}
where $G[A_i,J]$ is the functional containing the bulk integration.
The constraint (\ref{eq_33}) becomes $\partial_- A_i (x^-,\infty)=0$. 
Again, integrating over $x^-$ we have a restriction on field values at the infinity 
\begin{equation} \label{eq_39}
A_i (-\infty,\infty)=A_i (+\infty,\infty)
\end{equation}
Thus, we see that $A_i(\pm\infty,\vec x)$ are not independent variables.
More precisely, only a field $A_i$ obeying (\ref{eq_37}),(\ref{eq_33}), and (\ref{eq_39}) is an element of the phase space.

In principle, at the current stage, one might finish the determination of the phase space.
However, to construct a quantum theory, one should invert the symplectic form in the nonlinear manifold formed by
the field $A_i$ obeying the Gauss' constraints. 
Without additional assumptions, this is a difficult task.
Typically, the naive restriction $A_i(\infty)=0$ is used
or the more refined antisymmetric boundary condition $A_i(+\infty)=-A_i(-\infty)$ is adopted \cite{Zhang1,Zhang2}. 
However, we have seen that all such conditions are physically irrelevant. 
To obtain a tractable quantum theory having a complete physical information, 
we continue the further detailing of the phase space.

\subsection{Residual gauge fixing} \label{sub_5}
Although the residual gauge freedom is a well-know fact \cite{Zhang2}, it cannot be rigorously studied within the canonical method,
where, at first, one has to identify first class constraints.
Thus, the canonical method fails to reproduce a true phase space, since one has to use value of a field at the infinity as 
a canonical variable. In the symplectic method, we can safely use a constraint that involves field at a boundary.

The gauge fixing (\ref{eq_26}) is not complete. 
There exist gauge transformations that does not change $A_-$.
In fact, any $U(\vec x)$, which does not depend on $x^-$, is a symmetry of the dynamical system. 
So, after imposing the constraint (\ref{eq_26}), 
we again have the opportunity to impose an external constraint that does not miss physical information.  
To obtain a gain from a residual gauge fixing, we have to decide a constraint that will be used.
One might try to use the residual gauge freedom to make the antisymmetric boundary condition $A_i(-\infty,\vec x)=-A_i(+\infty,\vec x)$.
However, in general, this is impossible by the following reason.
Consider the transverse plane with two gauge connections $A_i(\vec x)$ and $B_i(\vec x)$. 
We are trying to find a gauge transformation $U(\vec x)$ that leads to $\tilde A_i=-\tilde B_i$.
Since $U(\vec x)$ must exist for any $A_i(\vec x)$ and $B_i(\vec x)$, it is sufficient to find a counterexample.
If $B_i=A_i$, then $\tilde A_i=\tilde B_i$ for any $U$. 
Hence, it is impossible to make an antisymmetric pair, until the case $F_{ij}[A]=0$ is in question.
Since a connection with zero curvature can be set to zero by gauge transformation, in this case the simple counterexample does not work.
The case of $F_{ij}=0$ will be processed later.

It is easily to show that the residual gauge freedom cannot make the symmetric boundary conditions, too.
Indeed, the condition $\tilde A_i=\tilde B_i$ and Eq. (\ref{eq_19}) immediately give the restriction $A_i=B_i$ 
which is obviously violated if $A_i$ and $B_i$ are arbitrary fields.

There is one more source of external constraints. We can restrict ourself to use only field configurations whose energy is finite.  
More precisely, we are free to impose constraints that remove only field configurations having infinite energy.
It is clear that, if $F_{ij}(\pm\infty,\vec x)\neq 0$, the Hamiltonian energy (\ref{eq_40}) is infinite.
So, we impose the new constraint  
\begin{equation} \label{eq_47}
F_{ij}(\pm\infty,\vec x)=0
\end{equation}
This means that at $x^-=\pm \infty$ the field $A_i(\vec x)$ must be a pure gauge field.
Although such fields have the trivial holonomy, it is again impossible to make the antisymmetric boundary condition 
from the residual gauge freedom. There exists an intrinsic algebraical obstruction. 
Let $A_i$ and $B_i$ be gauge fields. 
To make the antisymmetric boundary condition, one has to find a gauge transformation $U$ that leads to $\tilde A_i+\tilde B_i=0$, 
where transformed fields $\tilde A_i$ and $\tilde B_i$ was given in Eq. (\ref{eq_19}).
In the matrix notation, we have the equation
\begin{equation}
U (A_i+B_i) U^\dag - 2\frac{i}{g} (\partial_\mu U) U^\dag=0
\end{equation}
It has a solution only if the field $(A+B)/2$ is a pure gauge field. Hence, a solution exists only if
\begin{equation} \label{eq_42}
F_{ij}\left[\frac{A+B}{2}\right]=0
\end{equation}
Using the definition (\ref{eq_41}) and the constraints $F_{ij}[A]=F_{ij}[B]=0$, we can recast Eq. (\ref{eq_42}) to 
\begin{equation} \label{eq_43}
[A_i,A_j]+[B_i,B_j] - [B_i,A_j] - [A_i,B_j]=0
\end{equation}
To construct a counterexample, we choose $B_i=0$. In this choice, Eq. (\ref{eq_43}) becomes 
\begin{equation} \label{eq_44}
[A_i,A_j]=0
\end{equation}
If a theory has a non-Abelian gauge group and if it has two or more transverse dimensions, 
then there exists a pure gauge field that violates Eq. (\ref{eq_44}).
So, the inapplicability of the antisymmetric boundary condition has been proven.

Instead, we impose the following realizable constraint:
\begin{equation} \label{eq_45}
A_i(-\infty,\vec x)=0
\end{equation}
It is indeed realizable. Let $A_i(\vec x)=A_i(-\infty,\vec x)$ be a field such that $F_{ij}[A]=0$.
Let $\vec x_0$ be any point in the transverse plane. Let $U(\vec x,\vec x_0)$ be a gauge link from the point $\vec x_0$ to the point $\vec x$.
Since the field $A_i$ has zero curvature, $U(\vec x,\vec x_0)$ does not depend on form of a path, it only depends on the initial and final points of the paths.
To achieve the constraint (\ref{eq_45}), we perform the gauge transformation $V(\vec x)$
\begin{equation}\label{eq_46}
V(\vec x)=U^{-1}(\vec x, \vec x_0)=U(\vec x_0, \vec x)
\end{equation}
It is easy to check that after the gauge transformation (\ref{eq_46}) every gauge link becomes trivial. 
Also, it is important that there is no the Wu-Yang ambiguity \cite{Wu} which says 
that two distinct gauge potentials can have the same field strength. 
The strong constraint $F_{ij}=0$ completely removes such complications.  
So, we have proven that the gauge choice (\ref{eq_45}) is allowable and it completely fixes the residual gauge freedom. 

\subsection{The Hamiltonian} \label{sub_6}
Now we are ready to formulate the main result. 
At first, let us collect the obtained facts. 
Let $\gamma_i(\vec x)$ be value of a field $A_i(x^-,\vec x)$ at the boundary $x^-=+\infty$
\begin{equation}
\gamma_i(\vec x)=A_i(+\infty,\vec x)
\end{equation}
Due to Eq. (\ref{eq_47}) and (\ref{eq_37}),  $\gamma_i$ is not independent variable. It obeys the equations 
\begin{equation} \label{eq_48}
\partial_i \gamma_i^a(\vec x)= \int\limits_{-\infty}^{+\infty} \left(gf_{abc}\partial_-A_i^bA^c_i-gJ^+_a\right) dx^-
\end{equation}
\begin{equation} \label{eq_50}
\partial_i  \gamma_j^a-\partial_j  \gamma_i^a+g f_{abc} \gamma^b_i \gamma^c_j=0
\end{equation}
Moreover, Eqs. (\ref{eq_33}) and (\ref{eq_45}) give the asymptotic of $\gamma_i$ and $A_i$ at the transverse infinity
\begin{equation} \label{eq_58}
A_i(x^-,\infty)=0
\end{equation}
Note that the term $f_{abc}\partial_-A_i^bA^c_i$ in (\ref{eq_48}) is just the gluon contribution to $x^+$ component of the overall color current.
Thus, the constraint (\ref{eq_48}) has the form $\partial_i \gamma_i^a(\vec x)=g\rho^a(\vec x)$ 
where $\rho^a(\vec x)$ is the transverse density of color charge.

Now, we apply the method which was presented in Sec. \ref{sect_2}.
After the choice of the coordinates
\begin{equation} \label{eq_63}
\tilde A_i(x^-,\vec x)= A_i(x^-,\vec x) - \frac{1}{2}\gamma_i(\vec x)
\end{equation} 
we have
\begin{equation} \label{eq_64}
\tilde A_i(+\infty,\vec x)=-\tilde A_i(-\infty,\vec x)
\end{equation}
\begin{equation} \label{eq_49}
\tilde A_i(+\infty,\vec x)=\frac{1}{2}\gamma_i(\vec x)
\end{equation}
\begin{equation} \label{eq_51}
\partial_i \gamma_i^a(\vec x)= \frac{g}{2}f_{abc}\gamma_i^b\gamma_i^c+ \int\limits_{-\infty}^{+\infty} \left(gf_{abc}\partial_-\tilde A_i^b \tilde A^c_i-gJ^+_a\right) dx^-
\end{equation}
\begin{equation} \label{eq_52}
H=\frac{1}{2} (\pi^-_a)^2+\frac{1}{4}F_{ij}^a F^{ij}_a  - g A_i^a J^i_a
\end{equation}
\begin{equation} \label{eq_61}
\partial_-\pi^-_a + \partial_- \partial_i A_i^a-gf_{abc}\partial_-A_i^bA^c_i+gJ^+_a=0
\end{equation}
\begin{equation} \label{eq_53}
\omega= \int \partial_- d\tilde A_i^a \wedge d\tilde A_i^a
\end{equation}
We shall treat Eq. (\ref{eq_49}) as \emph{a constraint} that is defined in the space of pairs $(\tilde A_i,\gamma_i)$.
In this space, the constraints (\ref{eq_50}) and (\ref{eq_51}) completely fix the field $\gamma_i$ in terms of the field $\tilde A_i$.
Note, that the Hamiltonian (\ref{eq_52}) is uniquely defined only on the submanifold that is formed by the three constraints 
(\ref{eq_50}),(\ref{eq_49}), and (\ref{eq_51}).

\subsection{Quantization} \label{sub_7}
Unfortunately, the attempt to directly quantize the theory fails due to the lack of a set of coordinates that 
forms a linear space and solves the constraints.
In Sec. \ref{sect_2} we saw that if a phase space, having a continuous number of degrees of freedom, can be parameterized by  
a linear space, than the symplectic procedure can be safely applied
(a similar parametrization is just the isomorphism $\chi$ in Eq. (\ref{eq_12})).
So, to quantize the theory we have two distinct ways: either to solve the constraint or to introduce unphysical degrees of freedom.
We adopt the second way.  

If we remove the constraint (\ref{eq_49}), the phase space can be linearly parameterized by the set of fields $\tilde A_i$,
since the remaining constraints fix only the field $\gamma_i$.
The symplectic form (\ref{eq_53}) is already defined in the proper subspace.
After the fixation of $\gamma_i$, the Hamiltonian $H$ becomes a complicated functional over the field $\tilde A_i$.
Moreover, one can add to $H$ any functional that vanishes on the constraint surface (\ref{eq_49}).
The symplectic form $\omega$ is a nondegenerate form both with the constraint (\ref{eq_49}) and without it.

Now, the quantization is straightforward. 
The phase space is just tensor products of the phase space of the scalar field theory for each transverse index $i$ and each color adjoint index $a$.
Applying the rules (\ref{eq_54}--\ref{eq_55}), we can construct the Hilbert space in the basis of the free theory
which is the limit $g=0$.
Also, in the free theory, there are no unphysical degrees of freedom, since it is easy to check that $\gamma_i=0$ and
the constraint (\ref{eq_49}) gives $\tilde A_i(\infty,\vec x)=0$.
Dropping the constraint (\ref{eq_49}), we extend the phase space from the zero boundary condition to the antisymmetric boundary condition.
At lightcone, it is believed that these two boundary conditions give the equivalent quantum theory 
and one can safely use the extended phase space.  
Indeed, for the both cases the commutation relation (\ref{eq_54}) does not change.
However, in general, the quantum equivalence is not proven. 
Although it is well known that, in comparison with the free case, the complete Hamiltonian with an interaction can significantly change the structure of the Hilbert space,
we hope that for practical purposes of high energy evolution in QCD and the perturbation theory it is sufficient to work in the free Hilbert space
where possible unphysical degrees of freedom do not affect observables.

As we have seen, the symplectic method works well for linear spaces. 
In a more general case, to quantize a theory, a self-consistent Poisson algebra should constructed.
In addition, if a second class constraint exists, then the Dirac bracket should be found to obtain new constraintless Poisson algebra.
Indeed, from Poisson viewpoint the constraint (\ref{eq_49}) is a second class constraint.
Finally, the algebra should be quantized and appropriate Hilbert space with a Hamiltonian bounded from below should be constructed.
The problem is complicated by the existence of functionals involving spatial derivatives and nontrivial boundary conditions. 
In this paper, we just conjecture that the constraint (\ref{eq_49}) can be safely removed.
 
\subsection{Perturbative expansion} \label{sub_8}
The Hamiltonian (\ref{eq_52}) has the very complicated structure as a functional over the fundamental fields $\tilde A_i$.
Moreover, it contains arbitrary large power of the coupling constant $g$.
Indeed, $\gamma_i$, having finite order of $g$, cannot be a solution of Eq. (\ref{eq_50}).
For purposes of the perturbation theory it is useful to expand $H$ in the series on $g$.
Here we limit ourself to the first-order calculations 
and assume that the typical parton number has order $1$ (in a dense case it can have order $g^{-1}$).

In the zero-order $\gamma_i=0$ and we just obtain the free Hamiltonian (\ref{eq_56}).
Using Eq. (\ref{eq_63}), in the first-order the constraints (\ref{eq_50}) and (\ref{eq_51}) give
\begin{equation} \label{eq_57}
\partial_i  \gamma_j^a-\partial_j  \gamma_i^a=0
\end{equation}
\begin{equation} \label{eq_59}
\partial_i \gamma_i^a(\vec x)= \int\limits_{-\infty}^{+\infty} \left(gf_{abc}\partial_-\tilde A_i^b \tilde A^c_i-gJ^+_a\right) dx^-
=g\rho^a(\vec x)
\end{equation}
Eq. (\ref{eq_57}) says that the 1-from $\gamma_i dx^i$ is closed.
Since the transverse plane with the boundary conditions (\ref{eq_58}) has trivial de Rham 1-cohomology,
any closed 1-form is an exact form. This means that there exists 0-form $\varphi$ such that
\begin{equation}
\gamma_i=\partial_i \varphi
\end{equation}  
Hence, the solution of Eq. (\ref{eq_59}) is
\begin{equation} \label{eq_60}
\gamma_i^a(\vec x)= g\partial_i \frac{1}{\partial^2}\rho^a(\vec x)=\frac{g}{2\pi}\int \frac{x^i-y^i}{(\vec x - \vec y)^2} \rho^a(\vec y) dy
\end{equation}
where the integral representation has been written for two transverse dimensions.  
In the first order Eq. (\ref{eq_61}) becomes
\begin{equation} \label{eq_62}
\partial_-\pi^-_a + \partial_- \partial_i \tilde A_i^a+gj^+_a=0
\end{equation}
where we have defined the density of the overall color current 
\begin{equation} \label{eq_75}
j^+_a(x^-,\vec x)=-f_{abc}\partial_-\tilde A_i^b\tilde A^c_i+J^+_a
\end{equation}
which obviously obeys $\int j^+_a(\vec x,x^-) dx^-=\rho_a(\vec x)$.
To solve Eq. (\ref{eq_62}), we define the operator $\partial_-^{-1}$ by the following way:
\begin{equation} \label{eq_67}
\frac{1}{\partial_-}f(x^-)=\frac{1}{2}\int\limits_{-\infty}^{+\infty} \varepsilon(x^--y^-) f(y^-) dy^-
\end{equation}
Other choices of $\partial_-^{-1}$ are also allowed, but all they will give the same result.
Recalling the constraint (\ref{eq_32}) and the boundary condition (\ref{eq_64}), 
we are able to write the action of $\partial_-^{-1}$ on Eq. (\ref{eq_62})
\begin{equation} \label{eq_65}
\pi^-_a + \partial_i \tilde A_i^a+g\frac{1}{\partial_-}j^+_a=0
\end{equation}
Substituting $\pi^-_a$ from (\ref{eq_65}) into the Hamiltonian (\ref{eq_52})
and again using Eq. (\ref{eq_50}), in the first-order we have
\begin{equation} \label{eq_66}
H= H_0+ g \partial_i \tilde A_i^a \frac{1}{\partial_-}j^+_a + g \tilde A_i^a j^a_i
\end{equation}
where $j_i$ is the transverse components of the overall color current
\begin{equation}
j_i^a(x^-, \vec x)=-f_{abc} \partial_i \tilde A^b_j  \tilde A^c_j + J^a_i
\end{equation}

Note that in the derivation of the Hamiltonian (\ref{eq_66}) we did not use the solution (\ref{eq_60}) for $\gamma_i$.
In fact, in the first-order it is not required. Nevertheless, it will be required in the expansion up to the second-order.
Second-order calculations is more complicated and lies beyond the scope of this paper. 
For this case we only briefly sketch the calculations. Let $\xi_i$ be the second-order correction to $\gamma_i$.
Fortunately, before the third-order, Eqs. (\ref{eq_57}),(\ref{eq_59}), and (\ref{eq_60}) is still valid also for $\xi_i$. 
Eq. (\ref{eq_62}) is also valid, but instead of Eq. (\ref{eq_75}) we have
\begin{equation} 
j^+_a(x^-,\vec x)=-f_{abc}\partial_-\tilde A_i^b\tilde A^c_i-f_{abc}\partial_-\tilde A_i^b \gamma^c_i +J^+_a
\end{equation}
Finally, one can substitute $\pi^-$ from (\ref{eq_65}) into the Hamiltonian (\ref{eq_52}) and expand the result up to the second-order of the coupling constant $g$.
However, the explicit form is tedious and it is preferable to filter simultaneously the terms in according to a matrix element in question.

\subsection{Third order} \label{sub_9}
The expansion up to the third order is much more complicated, but it allows us to see the announced  
peculiarity of four space-time dimensions. Let $\eta_i$ be the third-order correction to $\gamma_i$.
The constraint (\ref{eq_50}) gives
\begin{equation} \label{eq_76}
\partial_i  \eta_j^a-\partial_j  \eta_i^a=-g f_{abc} \gamma^b_i \gamma^c_j
\end{equation}
Now, the 1-form $\eta=\eta_i dx^i$ is not closed. 
Using the language of differential form, we can rewrite Eq. (\ref{eq_76}) in the very compact form  
\begin{equation} \label{eq_77}
d\eta_a=\sigma_a
\end{equation}
where $\sigma_a$ is the given 2-form: $\sigma_a=-g f_{abc} \gamma^b_i \gamma^c_jdx^i \wedge dx^j$.
Eq. (\ref{eq_77}) immediately gives the integrability condition
\begin{equation} \label{eq_78}
d\sigma_a=0
\end{equation}
It is remarkable that in \emph{two transverse dimensions} the condition (\ref{eq_78}) is fulfilled automatically! 
If there are more than two transverse dimensions, then this condition must be checked explicitly.
If it is not fulfilled, then Eq. (\ref{eq_76}) does not have a solution and the whole theory fails to be consistent!
More precisely, in this case the finite-energy condition (\ref{eq_47}) cannot be imposed and the theory should be constructed in a different manner.
We shall not touch this problem and just assume $d\sigma_a=0$.

In the general case, the solution of (\ref{eq_76}) can be formulated with the help of the Poincar\'e duality.
There exists the Hodge map * that maps from $n$-forms to $(D-n)$-forms, where $D$ is the dimension of the transverse plane.
The Hodge map defines the Hodge adjoint of the exterior derivative: $\delta \sim *d*$.
Since the 1-cohomology is zero, the harmonic 1-forms do not exist and the Hodge theorem states that any 1-form is a sum of exact and coexact 1-forms
\begin{equation} \label{eq_79}
\eta=d \psi + \delta \theta
\end{equation}
where $\psi$ is a 0-form and $\theta$ is a 2-form.
The exact part $d \psi$ disappears from (\ref{eq_77}) as well as in the first order.
So, Eq. (\ref{eq_77}) becomes
\begin{equation}
d\delta \theta=\sigma
\end{equation}
If $d\sigma_a=0$ then a solution for $\theta$ obviously exists.
The coexact form $\delta \theta$  disappears from Eq. (\ref{eq_59}), since it is easily to prove the identity  $\partial_i (\delta \theta)_i$.
So, $\partial_i \psi$ can be directly found from the extension of Eq. (\ref{eq_60}) where now $\rho^a(\vec x)$ is a more complex function over $\tilde A_i$,$J^+$,$\gamma_i$, $\xi_i$.

If the transverse plane is two-dimensional, the Hodge decomposition (\ref{eq_79}) has the more simple form
\begin{equation}
\eta_i=\partial_i \psi + \varepsilon_{ji} \partial_j \bar  \psi
\end{equation}
where $\psi(\vec x)$ and $\bar  \psi(\vec x)$ are two arbitrary functions.
The tensor $\varepsilon_{ij}$ is the alternating tensor such that $\varepsilon_{ij}=-\varepsilon_{ji}$ and $\varepsilon_{12}=1$.
The constraints (\ref{eq_50}) and (\ref{eq_51}) uniquely determine $\psi$ and $\bar  \psi$ via the following equations:
\begin{equation}
\Delta \psi = F_1
\end{equation}
\begin{equation}
\Delta \bar \psi = F_2
\end{equation}
where $\Delta=\partial_i\partial_i$ is the Laplace operator and
$F_1$, $F_2$ is functions depending on $\tilde A_i$,$J^+$,$\gamma_i$, $\xi_i$.

The fourth and higher orders can we processed in the way that has been developed for the third order.

\section{JIMWLK at once} \label{sect_6}
In this section, to check our result, we straightforwardly derive the JIMWLK equation \cite{JIMWLK} from the first-order Hamiltonian. 
The general method of derivation was developed in Ref. \cite{avp_JIMWLK}.
It is argued that only one soft gluon emission into new opened phase space contributes to the evolution of the elastic scattering amplitude,
since only a linear term over small $\delta Y$ survives in the differential equation.
Although it is proved that one needs to known boosted wave function of a projectile with second order accuracy of the Hamiltonian  
perturbation theory, the key building block is the soft gluon emission amplitude of the first-order.

In Ref. \cite{avp_JIMWLK} the different gauge condition is used. 
To use the Hamiltonian (\ref{eq_66}), we have to recalculate the matrix element of the interaction part.
Only the term $\partial_i \tilde A_i^a \partial_-^{-1}j^+_a$ gives a contribution to this matrix element.
The operator $\partial_i \tilde A_i^a$ emits a soft gluon, while the operator $j^+_a$ probes the fast color charges in the projectile.
In the first-order, the wave function of the boosted projectile with a soft gluon can be calculated with the help of Eq. (21) in \cite{avp_JIMWLK}:
\begin{equation} \label{eq_71}
\ket {P^{(1)}}=-H^{-1}_0 V  \ket {P^{(0)}}
\end{equation}
where $V=\int \partial_i \tilde A_i^a \partial_-^{-1}j^+_a$.
Using the definition (\ref{eq_67}) and the gauge version of (\ref{eq_68}), we need to calculate the integral
\begin{equation} \label{eq_69}
\int e^{ik^+x^-} \frac{1}{2}\varepsilon(x^--y^-) dx^-= e^{ik^+y^-} \int e^{ik^+x^-} \frac{1}{2}\varepsilon(x^-) dx^-
\end{equation}
where $k^+$ is the longitudinal momentum of the soft gluon.
The integral in Eq. (\ref{eq_69}) can be calculated with the help of the following integral identity for any $t>0$: 
\begin{equation}
\int\limits_{-\infty}^{+\infty}e^{itx}\frac{1}{2}\varepsilon(x)dx=\frac{i}{t}
\end{equation}
Then, we obtain the integral
\begin{equation} \label{eq_70}
\int e^{ik^+y^-} j^+_a(y^-,\vec x) dy^-
\end{equation}
Since the operator $j^+_a$ probes the color charges having a high longitudinal momentum, in the evaluation of the integral in Eq. (\ref{eq_70})
we can neglect the small longitudinal momentum $k^+$ of the soft gluon. 
Hence, the integral (\ref{eq_70}) can be expressed as
\begin{equation}
\left. \int e^{ik^+y^-} j^+_a(y^-,\vec x) dy^- \right|_{k^+=0}=  \int j^+_a(y^-,\vec x) dy^-  = \rho_a(\vec x)
\end{equation}
The action of the operator $H^{-1}_0$ is $2k^+/\vec k^2$ as it is shown in \cite{avp_JIMWLK}. 
Collecting all terms we obtain from Eq. (\ref{eq_71}) the first-order correction to the projectile wave function
\begin{equation} \label{eq_73}
\ket {P^{(1)}}= g \int \frac{2k_i}{\vec k^2} \frac{e^{-i\vec k \vec x}}{\sqrt{2k^+}\sqrt{(2\pi)^3}} 
\hat a^\dag_{\vec k,k^+,i,a} \hat \rho^a(\vec x) \ket {P^{(0)}} d^2\vec x d^2\vec k dk^+
\end{equation}
where the integration over $k^+$ is performed in the new opened rapidity window.
It is useful to define the gluon creation operator in a given transverse point
\begin{equation} \label{eq_74}
 \hat{a}^\dag_{\vec{x},k^+,i,a}=\int e^{-i\vec{k}\vec{x}} \hat{a}^\dag_{ \vec k,k^+,i,a} \frac{d^2k}{\sqrt{(2\pi)^2}}
\end{equation}
To express Eq. (\ref{eq_73}) in the term of Eq. (\ref{eq_74}), we have to calculate the following integral:
\begin{equation}
\int \frac{\vec k}{k^2}  e^{i\vec k\vec r}d^2\vec k=2\pi i\frac{\vec r}{r^2}
\end{equation}
Finally, at the first-order we have
\begin{equation}  \label{eq_72}
 \ket{P^{(1)}} = \int \frac{ig}{2\pi\sqrt \pi \sqrt{k^+}}   \hat a^\dag_{\vec y,k^+,i,a}  \frac{(\vec y-\vec x)_i }{(\vec y-\vec x)^2} 
 \hat\rho^a(\vec x)\ket{P^{(0)}} dk^+ d^2\vec x d^2\vec y
\end{equation}  
Eq. (\ref{eq_72}) has a clear physical description. 
A fast color source at a transverse point $\vec x$ can emit the soft gluon into arbitrary point $\vec y$ of the transverse plane.

The further calculation is fully equivalent to Ref. \cite{avp_JIMWLK}. 
If the projectile scatters in the target color field $\alpha_a(\vec x)$, 
the evolution equation for the elastic scattering amplitude is
 \begin{equation} 
 \frac{dS[\alpha]}{dY}=\frac{g^2}{(2\pi)^3}\int\limits_{zxy}^{\phantom{z}} K_{zxy}
 \bra P 
 \left(
 \begin{array}{c}
 - \hat S \hat \rho^a(\vec y)  \hat \rho^a(\vec x) \\
 - \hat \rho^a(\vec y) \hat \rho^a(\vec x) \hat S \\
+2V_{ba}(\vec z) \hat \rho^b(\vec y) \hat S \hat \rho^a(\vec x) 
 \end{array}   \right)
\ket P 
 \end{equation}
\begin{equation}
 K_{zxy}=\frac{(\vec z-\vec y)(\vec z-\vec x)}{(\vec z-\vec y)^2(\vec z-\vec x)^2}
\end{equation}
where $V_{ba}(\vec z)$ is the gluon scattering amplitude which is a Wilson line in the adjoint representation. 
The operator $\hat S$ is the operator of $S$-matrix.

Let us briefly consider the general properties of high-order corrections. The detailed study will be a subject of a separate paper. 
It should be emphasized that we assume that the number of partons in the projectile has the order $O(1)$.
This differs from the approach of Ref. \cite{Kovner07} where the order $O(1/g)$ was assumed.
This distinction gives different corrections to the JIMWLK equation. 
Thus, a physical meaning of each approach should be clarified.

The careful examination of the terms in the Hamiltonian gives the following form for an operator 
that has a nonvanishing contribution to the JIMWLK equation:
\begin{equation} \label{eq_80}
\tilde A_i \left(\frac{1}{\partial_-} j^+_a\right) D[j^+_a] 
\end{equation}
where $D[j^+]$ is a polynomial over the current $j^+_a$. 
Indeed, the operator $A_i$ creates only one soft gluon, while all other operators must act on the fast partons.
Only the operator $j^+_a$ gives nonvanishing matrix element in the high-energy limit.
The integral kernel $\partial_-^{-1}$ is required to compensate the multiplier $k^+$ in the numerator, 
which comes from $H^{-1}_0$ in the perturbation theory (\ref{eq_71}).  The convolution of the space and color indexes in (\ref{eq_80}) is not restricted. 

Analyzing the Hamiltonian (\ref{eq_52}), we conclude that the form (\ref{eq_80}) arises only from the term $(\pi^-_a)^2$.
In the first order we already used the corresponding term $\partial_i \tilde A_i^a \partial_-^{-1}j^+_a$.
In the second order there are no terms having the form (\ref{eq_80})!
In the third order only the following term gives a contribution to the JIMWLK equation:
\begin{equation}
V=-g^2 \tilde A_i^b \gamma_i^c f_{abc} \frac{1}{\partial_-} j^+_a
\end{equation}
where $\gamma_i$ is given by the first-order expression (\ref{eq_60}).
The calculation of the wave function of the boosted projectile is very similar to the first-order calculation.
Starting from Eq. (\ref{eq_71}), we obtain the following correction to the wave function
\begin{equation} \label{eq_81}
\ket{P^{(1)}_3}= \int
\frac{ig^3f_{abc}}{8\sqrt{\pi^3k^+}}
\frac{(\vec x_1-\vec x_2)^i}{(\vec x_1-\vec x_2)^2} \ln \frac{|\vec z-\vec x_1||\vec z-\vec x_2|}{\mu^2}
\hat a^\dag_{\vec z,k^+,i,b} :\hat\rho^c(\vec x_1) \hat\rho^a(\vec x_2): \ket{P^{(0)}}
\end{equation} 
where $\mu$ is an infrared regulator and the integration is over $\vec x_1,\vec x_2,\vec z,k^+$. 
Note that in higher orders there exists the problem of operator ordering in the Hamiltonian.
We adopt here the standard prescription: it is assumed that the Hamiltonian is already normal ordered $H\equiv :H:$.
From practical viewpoint, such normal ordering means that two operators $\hat\rho^a(\vec x_1)$ and $\hat\rho^c(\vec x_2)$ cannot act on a same parton.
Thus, in Eq. (\ref{eq_81}) the normal ordering is required.
Applying the algorithm of Ref. \cite{avp_JIMWLK}, we obtain the following $O(\alpha_s^2)$-correction to the JIMWLK equation: 
 \begin{equation} 
 \frac{dS_2[\alpha]}{dY}=\frac{g^4}{32\pi^3}\int K_{zx_1x_2y}
 \bra P 
 \left(
 \begin{array}{c}
 - \left\{ \hat S, \left\{\hat \rho^a(\vec y), \hat \tau^a(\vec x_1,\vec x_2)\right\}\right\} \\
+2V_{ba}(\vec z) \hat \rho^b(\vec y) \hat S \hat \tau^a(\vec x_1,\vec x_2)  \\
+2V_{ba}(\vec z) \hat \tau^b(\vec x_1,\vec x_2) \hat S \hat\rho^a(\vec y)
 \end{array}   \right)
\ket P 
 \end{equation}
\begin{equation}
K_{zx_1x_2y}= \frac{(\vec z-\vec y)(\vec x_1-\vec x_2)}{(\vec z-\vec y)^2(\vec x_1-\vec x_2)^2} \ln \frac{|\vec z-\vec x_1||\vec z-\vec x_2|}{\mu^2}
\end{equation}
\begin{equation} \label{eq_82}
\hat \tau^b(\vec x_1,\vec x_2)=f_{abc} :\hat\rho^c(\vec x_1) \hat\rho^a(\vec x_2):
\end{equation}
where the bracket $\{,\}$ denotes the anticommutator. 
Note that Eq. (\ref{eq_81}) does not give the $O(\alpha_s^3)$-correction to the JIMWLK equation,
because, in addition, one must also examine the $O(g^5)$-terms in the Hamiltonian.

\section{Conclusion} \label{sect_4}
In this paper, we have constructed the complete lightcone Hamiltonian of QCD. 
It has the very complicated structure and infinite power over the coupling constant $g$.
However, the perturbative expansion can be performed in a controlled way.
A further task is to apply the developed method to study of various physical problem.
Specifically, one can systematically calculate higher corrections to the JIMWLK equation (sometimes, they are called NLO, NNLO and so on).  

Unexpectedly, our calculations give a restriction on the dimension of the space-time.
If the space-time is three-dimensional, the residual gauge freedom can be directly used to make the antisymmetry boundary condition.
Indeed, in one transverse dimension Eq. (\ref{eq_42}) holds automatically. Hence, there is no reason to introduce boundary field $\gamma_i$.
So, three-dimensional QCD has the much more simple Hamiltonian which contains a finite number of terms and a finite power of $g$.
It is natural to expect that this theory is too insignificant in comparison with the four-dimensional QCD.

If the space-time has more than four dimensions, a quantization of the theory should be different. 
Probably, the finite-energy condition (\ref{eq_47}) has no solutions. 
This means that all excitations of such a theory have infinite energy.
Although now we cannot prove that higher-dimensional QCD does not exist, the exceptional role of four-dimensional QCD is clearly visible.
This result is not very surprising. 
The history of quantum field theory knows enough examples where the theory chooses a geometry.   

For completeness, the two-dimensional QCD should be mentioned.  
The developed method does not give any profit in this case.
The reasons are: 
the field $A_i$ does not exist, 
the residual gauge freedom degenerates to global color rotations,
there are no propagating degrees of freedom -- there are no gluons,
and obviously high energy evolution is trivial.
However, the theory is not fully trivial -- it is still possible to study the $q\bar q$-potential and Wilson loops.

Let us make several remarks about the method of lightcone coordinates.
Although after the first studies this method rises great expectancies, 
today it is clear that the method is more complicated. 
While in the Cartesian coordinates the complexity of the theory is pushed entirely into the quantum level,
in the lightcone coordinates it remains in the classical level via the complicated structure of the phase space and the uncomputable Hamiltonian.
Due to the ``conservation'' of the complexity, there are no reasons to expect a simple solution in the both approaches. 
However, in the lightcone method the complexity grows in a controlled manner.
Indeed, in the classical level we work with the classical mathematical background which is much more friendly than the Hilbert space.
Possibly, since the lightcone theory has the trivial vacuum, the solution of many problems can be found in the classical level.  
Since the Lagrangian can be generalized to a fully covariant form, 
there is no a priori frames to quantize the theory. 
Hence, the lightcone coordinates is a natural choice as well as a choice of the Cartesian coordinates.

\section*{Acknowledgments}
I thank N.V. Prikhodko for the reading of the manuscript.

\end{document}